\documentclass [aps,pra,amssymb,amsmath,twocolumn,showpacs]{revtex4-1}

\usepackage{amsmath}
\usepackage{amssymb}
\usepackage{graphicx}
\usepackage{verbatim}
\usepackage{bm}
\usepackage{color}

\usepackage{graphicx,epsfig,amsfonts,amssymb}
\usepackage{bm}
\usepackage{times}
\usepackage{lipsum}

\usepackage[all]{xy}

\newcommand{\nn}{\nonumber}

\newcommand{\ra}{\rangle}
\newcommand{\rar}{\rightarrow}
\newcommand{\da}{\downarrow}
\newcommand{\ua}{\uparrow}
\newcommand{\be}{\begin{eqnarray}}
\newcommand{\ee}{\end{eqnarray}}

\newcommand{\bs}{\begin{equation}\begin{split}}
\newcommand{\es}{\end{split}\end{equation}}

\date{\today}

\begin{document}
\title{Dynamic spin localization and $\gamma$-magnets}

\author{Vladimir Y. Chernyak$^{a,b}$, Nikolai A. Sinitsyn$^c$, Chen Sun$^{d}$}
\affiliation{$^a$Department of Chemistry, Wayne State University, 5101 Cass Ave, Detroit, Michigan 48202, USA}
\affiliation{$^b$Department of Mathematics, Wayne State University, 656 W. Kirby, Detroit, Michigan 48202, USA}
\affiliation{$^c$Theoretical Division, Los Alamos National Laboratory, Los Alamos, NM 87545, USA}
\affiliation{$^d$Department of Physics, Brown University, Providence, Rhode Island 02912, USA}


\begin{abstract}
We  explore an unusual type of quantum  matter that can be realized by qubits  having different physical origin.   Interactions  in this matter are described by essentially different coupling operators  for all qubits. 
We show that at least  the simplest such  models, which
can be realized with localized states in Dirac materials,
  satisfy integrability conditions that we use to describe pseudospin dynamics in a linearly time-dependent magnetic field.
 Generalizing to an arbitrary number of qubits, we construct a spin Hamiltonian, which we call the $\gamma$-magnet. This system does not conserve  polarization of any spin and the net spin polarization. Nevertheless, for arbitrarily strong interactions, nonadiabatic dynamics,  and any initial eigenstate,  we find that  quantum interference  suppresses spin-flips. 
This behavior resembles many-body localization but occurs in  phase space of many spins rather than real space. This effect may not have a counterpart in classical physics and can be a signature of a new type of  spin ordering, which is different from both disordered spin glasses and ordered phases of spin lattices.
\end{abstract}

\maketitle
\section{Introduction}
In addition to spins, electrons in solid state generally carry additional pseudo-spins: valley and band indices, position inside two coupled quantum dots, e.t.c.. Every such a discrete index can add an independent qubit for quantum information-processing. Hence, it is possible to perform basic quantum computations using just one electron, without the need to entangle different qubit carriers. Moreover, qubits of different physical origin can be conveniently  accessible by different fields individually. For example, the true spin is coupled to the magnetic field whereas the energies of spatially separated electron states can be controlled by electric voltages.

Heterogeneous multi-qubit systems are different from the commonly known magnets with exchange or dipole interactions because not only the strength but also the type of interaction is different now for different qubits. Therefore, here we raise the main question of this article: can heterogeneous interacting qubits show  collective effects that are not found in conventional quantum spin models? 

This question would be hard to address by standard theoretical means that essentially rely on the assumption of identical, at least in the statistical sense, spin interactions. Here, we propose  a different approach by observing that some of the most elementary models of heterogenious qubits turn out to be integrable and generalizable to arbitrary number $N$ of interacting pseudo-spins. By deriving an exact solution for their behavior in a time-dependent magnetic field, we will show an unusual effect that we named {\it dynamic spin localization} (DSL), which means the ``yes" answer to our question. 

\section{Two qubits: Bound states of Dirac electrons}
 The discovery of graphene and 2D Dirac semiconductors open the possibility to assign simultaneously several pseudospins to a single electron, including the valley and Dirac sub-band indices, as well as the layer index in the case of multi-layer material. Additional electron confinement in a region with several metastable states can boost this number. The phase space of such a multi-qubit system may then be comparable to a mesoscopic spin cluster, e.g., a  nanomagnet.

 As the most trivial example, consider a weakly bound state near an impurity of a Dirac semiconductor \cite{Dirac}. Conduction electrons  carry both spin and valley degrees of freedom that are described by corresponding Pauli operators $s_{\alpha}$ and $\tau_{\alpha}$, where $\alpha=x,y,z$. A weakly bound state  then carries the same degrees of freedom, so the effective Hamiltonian of these two qubits in an external magnetic field contains the terms that are normally found in conduction electrons: 
 \be
 H_0 = \beta_{\tau} B \tau_z +\beta_s B s_z +\varepsilon \tau_z s_z,
 \ee
 where $\beta_s$ and $\beta_{\tau}$ are the effective g-factors that describe couplings to the external out-of-plane magnetic field $B$, and $\varepsilon$ is the  effect of the Kane-Mele spin-orbit coupling \cite{KM-graphene}. This Hamiltonian conserves qubit polarizations. However, a short range nonmagnetic impurity potential  mixes degenerate localized states from different valleys. This mixing is described by the effective coupling $\sim \tau_x$. In addition, the impurity potential may introduce its own intrinsic spin-orbit coupling. Since $\tau_z$ and $s_{\alpha}$ are odd under time-reversal, the additional allowed spin orbit coupling has the form $\sim \tau_z s_x$. Thus, the most general two-qubit Hamiltonian for such a localized state is given by
  \be
 H = \beta_{\tau} B \tau_z +\beta_s B s_z +\varepsilon \tau_z s_z +g_1\tau_x+g_2\tau_z s_x.
 \label{two-qubits}
 \ee   
 
 This Hamiltonian does not conserve qubit polarizations, and for arbitrary values of parameters there is no simple explicit expression for its eigenvalues. However, it is integrable in the sense that we can write a simple expression for the Hamiltonian that commutes with it for arbitrary values of all parameters:
  \be
 H' =  B \tau_z s_z +  \frac{\varepsilon}{\beta_{\tau} } \tau_z + \frac{\varepsilon}{\beta_s} s_z+ \frac{g_1}{\beta_\tau} \tau_x s_z+\frac{g_2}{\beta_s} s_x,
 \label{two-qubits2}
 \ee    
\be
[H,H']=0.
\label{comHH}
\ee 
Hamiltonians with such simple commuting partners attract lot of attention for their comprehensive dynamics during quantum quenches and  thermalization \cite{integrable-thermal,int1}.  They  often have  Poisson statistics of energy level splittings and frequent appearance of exact energy level crossings \cite{patra-LZ}. 

According to \cite{commute},  a  system with a  time-dependent Hamiltonian $H_1(t,\varepsilon)$, where $t$ is time and $\varepsilon$ is  a constant parameter, can be solvable if there is a nontrivial Hamiltonian $H_2(t,\varepsilon)$, such that the following two conditions are satisfied:
\begin{eqnarray}
\label{cond1}
&& [{H}_1(t,\varepsilon),H_2(t,\varepsilon)]=0, \\
\label{cond2}
&&\partial {H}_1/\partial \varepsilon =  \partial {H}_2/\partial t,\quad \quad \forall \, t,\varepsilon.
\end{eqnarray}
The pair $H$ and $H'$ does satisfy the relation
\be
\frac{\partial H}{\partial \varepsilon}= \frac{\partial H'}{\partial B}.
\label{partialH}
\ee
Hence, according to Ref.~\cite{commute}, dynamics with the Hamiltonian $H$ that satisfies (\ref{comHH}) and (\ref{partialH}) can be also understood analytically if $B$ is linearly time-dependent. The explicitly time-dependent Hamiltonian 
  \be
 H(t) = \beta_{\tau} t \tau_z +\beta_s t s_z +\varepsilon \tau_z s_z +g_1\tau_x+g_2\tau_z s_x
 \label{two-qubits3}
 \ee  
describes a situation that typically appears during the measurement of a hysteresis loop, so that the magnetic field has to  sweep from large negative to large positive values. 
In Dirac semiconductors, sufficiently strong and fast time-dependent effective magnetic fields can be induced optically \cite{four-LZ}. 
 
 As $t\rar \pm \infty$, qubits cannot flip due to the strong field along $z$. 
Thus, we can consider a solvable scattering problem such that qubits start as an eigenstate of $\tau_z$ and $s_z$ operators as $t\rar -\infty$. Such states are called diabatic states. The problem is to determine qubit polarizations as $t\rar +\infty$. This information is contained in the transition probability matrix $P$, whose element $P_{nm}$ is the probability that our system is at the diabatic state  $n$  at $t=+\infty$ given that at $t=-\infty$ the system is in the state $m$.

The solution of this problem for the model (\ref{two-qubits}) has been previously conjectured in \cite{four-LZ}, and the rigorous proof of this solution can be  found in \cite{commute}. 
The states
\be
|\ua \ua \ra, \,\,\, |\da \da \ra, \,\,\, |\da \ua \ra, \,\,\, |\ua \da \ra,
\label{eig-4}
\ee
are eigenstates of the Hamiltonian (\ref{two-qubits}) at $t=\pm \infty$. 
In the basis (\ref{eig-4}) the transition probability matrix is given by \cite{four-LZ}
\be
P^{N=2}=\left( \begin{array}{cccc}
p_1p_2 & 0 & q_1p_2 & q_2  \\
0 & p_1 p_2 & q_2 & q_1 p_2\\
q_1p_2 & q_2& p_1p_2 & 0\\
q_2 & q_1 p_2 & 0 & p_1p_2
\end{array} \right),  \quad  \beta_1<\beta_2,
\label{prob-sol2}
\ee
where
$$
p_{1,2}=e^{-\pi g_{1,2}^2/\beta_{1,2}}, \quad q_{1,2}=1-p_{1,2}.
$$

\section{Three qubits}
Apart from spin and valley, Dirac-like Hamiltonians generally contain other discrete degrees of freedom, such as electron and hole index of a Dirac cone. 
For example, the graphene Hamiltonian with Kane-Mele type of spin-orbit coupling is given by \cite{KM-graphene}
\be
H_{KM}=v(k_x\tau_z \sigma_x +k_y \sigma_y)+\varepsilon \sigma_z \tau_z s_z,
\label{km1}
\ee
where  $k_{x,y}$ are effective momenta of electrons, and Pauli operators $\sigma_{\alpha}$ originate from the difference of two sites in the unit cell of the honeycomb lattice of graphene. This degree of freedom  is time-reversal invariant, and does not couple to the magnetic field but it can couple to the other two pseudospins by spin orbit coupling. 

 A short range non-magnetic impurity can create bound states near  zero energy of the Hamiltonian $H_{KM}$. Qubits would then become additionally coupled by the  intrinsic to the impurity spin-orbit coupling. In addition to the Kane-Mele term with strength $\varepsilon$, time-reversal invariance allows then additional terms $\sim \tau_z s_z$, $\sim \tau_zs_x$, and $\sim \tau_z s_z \sigma_x$. So, a realistic model of qubit interactions for this localized state in a linearly time-dependent magnetic field has the Hamiltonian 
\begin{eqnarray}
\label{km2}
H=t \left[\beta_1 \tau_z +\beta_2 s_z \right] +\varepsilon \tau_zs_z\sigma_z &+& \\
\nonumber \varepsilon' \tau_zs_z +g_1 \tau_x + g_2 \tau_z s_x &+& g_3 \tau_zs_z \sigma_x,
\end{eqnarray}
where $\beta_{1,2}$, $\varepsilon'$, $\eta$, and $g_{1,2,3}$ are constant parameters. 

To verify that the model (\ref{km2}) is integrable, we searched for the Hamiltonians that would depend linearly on $t$ and $\varepsilon$ and satisfy relations (\ref{cond1}) and (\ref{cond2}). One of these Hamiltonians was chosen to contain the same types of qubit couplings as in Eq.~(\ref{km2}). We found that there is a family that contains not just two but three independent Hamiltonians:
\begin{eqnarray}
\label{famH1}
\nonumber H_1&=&\varepsilon \tau_zs_z\sigma_z+t [\beta_1 \tau_z+\beta_2s_z+\beta_3\sigma_z] + \\
\nonumber &+&\eta [\beta_1\beta_2 \tau_zs_z +\beta_1\beta_3\tau_z\sigma_z+\beta_2\beta_3s_z\sigma_z]  +\\
 &+&g_1\tau_x+g_2\tau_zs_x +g_3 \tau_zs_z\sigma_x,\\
\label{famH2}
\nonumber \\
\nonumber H_2&=&t\tau_zs_z\sigma_z +\varepsilon \left[\frac{\tau_z}{\beta_1} +\frac{s_z}{\beta_2}+\frac{\sigma_z}{\beta_3}\right]+\\
\nonumber  &+&\eta [\beta_3\tau_zs_z+\beta_2\tau_z\sigma_z+\beta_1s_z\sigma_z] +\\
&+&\frac{g_1}{\beta_1}\tau_xs_z\sigma_z +\frac{g_2}{\beta_2} s_x\sigma_z +\frac{g_3}{\beta_3}\sigma_x,\\
\label{famH3}
\nonumber \\
\nonumber H_3&=&\varepsilon [\beta_3\tau_zs_z+\beta_2\tau_z\sigma_z+\beta_1s_z\sigma_z]+\\
\nonumber&+&t[\beta_1\beta_2\tau_zs_z+\beta_1\beta_3\tau_z\sigma_z+\beta_2\beta_3s_z\sigma_z]+\\
\nonumber&+&\eta [2\beta_1\beta_2\beta_3 \tau_zs_z\sigma_z+\beta_1(\beta_2^2+\beta_3^2)\tau_z+\\
\nonumber &+&\beta_2(\beta_1^2+\beta_3^2)s_z + \beta_3(\beta_1^2+\beta_2^2)\sigma_z]+\\
\nonumber&+&g_1(\beta_3\tau_x\sigma_z+\beta_2 \tau_xs_z )+g_2(\beta_3 \tau_zs_x\sigma_z+\beta_1 s_x) +\\
&+&g_3(\beta_2\tau_z\sigma_x +\beta_1s_z\sigma_x).
\label{HHH1}
\end{eqnarray} 
In addition to commuting with each other, they satisfy the relations
\be
\frac{\partial H_1}{\partial \varepsilon}=\frac{\partial H_2}{\partial t}, \,\,\, \frac{\partial H_1}{\partial \eta}=\frac{\partial H_3}{\partial t}, \,\,\,\, \frac{\partial H_2}{\partial \eta}=\frac{\partial H_3}{\partial \varepsilon}.
\ee
The Hamiltonian (\ref{km2}) is a special case of this family. It is obtained from $H_1$ by setting $\beta_3=0$ and identifying $\varepsilon'=\eta \beta_1\beta_2$. So, the transition probability matrix for the model (\ref{km2}) can be obtained explicitly, as well as for any of the Hamiltonians $H_1$, $H_2$, and $H_3$. 

The class of possible integrable time-dependent Hamiltonians with only three spin-like variables is not restricted to the three Hamiltonians (\ref{famH1})-(\ref{famH3}). According to \cite{commute,large-class}, if we introduce constant parameters $a$, $b$, $c$, $e_1$, $e_2$, and $e_3$ then we can also introduce a new time variable $T$ such that 
\begin{eqnarray}
t\!=\!aT+e_1, \quad \varepsilon \!=\!bT+e_2, \quad \eta\!=\!cT +e_3.
\label{timeHHH}
\end{eqnarray} 
Then, the Hamiltonian 
\be
\nonumber H(T)\!=\!a H_1(T) \!+\!b H_2(T) \!+\!c H_3(T)
\label{largeHHH}
\ee
belongs to the class of solvable multistate Landau-Zener systems. 
The way to obtain an explicit expression for  the transition probability matrix for evolution with such Hamiltonians is described in detail in \cite{large-class,quest-LZ}. 

Moreover, the family (\ref{famH1})-(\ref{famH3}) is only a special case of integrable families called multitime Landau-Zener models \cite{large-class}. There are also possibilities to construct integrable interacting spin families that contain operators with, e.g., $\sim 1/t$ time-dependence of some of the parameters \cite{yuzbashyan,bcs}. 
 Hence, the possibilities to find solvable time-dependent systems are 
numerous. Therefore, quantum integrability is an attractive possibility to study behavior of heterogeneous interacting qubits. This is not only because  solvable  models provide  the insight into  nonperturbative regime but also because they can do this when standard methods, which usually deal with identical spin types, cannot be used.

\section{Gamma magnets}
\begin{figure}[!htb]
\scalebox{0.5}[0.5]{\includegraphics{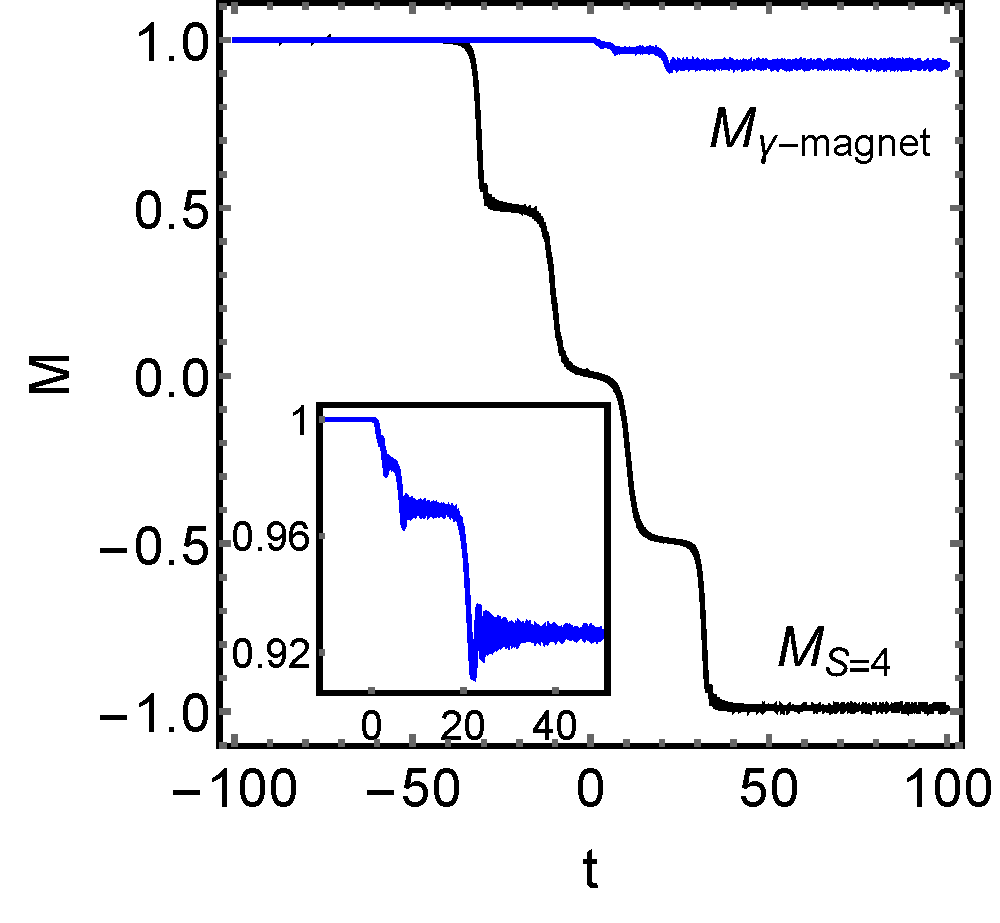}}
\caption{Numerically found normalized  average magnetization of   $S\!=\!4$ spin with the Hamiltonian $H(t)\!=\!-a S_z^2 \!+\! b S_x^2 \!+\! \beta t S_z $ (black curve;  $a = 5$, $b = 0.5$, $\beta= 1$), and a  $\gamma$-magnet (\ref{ham-string1})  with $N\!=\!5$ (blue curve and inset; $\varepsilon\!=\!-10$, $\beta_1\!=\!0.5$, $\beta_2\!=\!1.7$, $\beta_3\!=\!4.1$, $\beta_4\!=\!7.1$, $\beta_5\!=\!9.2$, $g_1\!=\!g_5\!=\!0.14$, $g_2\!=\!g_4\!=\!0.15$, $g_3\!=\!0.17$). Evolution starts from the highest energy state. }
\label{gamma-FE-magnetiztion-compare}
\end{figure}

The unusual behavior of heterogeneous qubit systems is fully revealed if we look at their dynamics for arbitrary number $N$ of interacting qubits.  
We will demonstrate  a quantum interference effect that occurs with quantum spins in a linearly changing external field. Interacting spin clusters usually
reverse their magnetization after passing through several resonances in a slow linearly time-dependent field, as the black magnetization curve   in Fig.~\ref{gamma-FE-magnetiztion-compare} shows for spin $S=4$ nanomagnet with quadratic anisotropy. 
At least in the adiabatic limit, the magnetization usually follows the direction of the external field, which changes sign during the hysteresis measurements.
We will show that heterogeneous spin interactions can make the magnetization in a linearly time-dependent field behave in a  radically different way. 

After the field changes with any rate between strongly different values, the state of our system  ends up close to the initial state, despite spin not being conserved, as we show by the blue curve for the  magnetization  in Fig.~\ref{gamma-FE-magnetiztion-compare}.
This happens for any multi-spin interaction strength and any initial state.  This is what we call DSL.
We will demonstrate this effect by constructing an analytically solvable model that shows it.

To construct an integrable $N$-spin Hamiltonian, let us return to the two-qubit one, in Eq.~(\ref{two-qubits}), and note that 
its coupling terms 
\be
\gamma_1\equiv \tau_x, \quad {\rm and} \quad \gamma_2 \equiv \tau_zs_x
\label{two-g}
\ee
are represented by 4$\times$4 matrices that satisfy the relations: 
\be
\{\gamma_1,\gamma_2\} =0, \quad \gamma_1^2=\gamma_2^2 = 1.
\label{two-g1}
\ee
Thus, the coupling terms are Dirac $\gamma$-matrices in 4D space. There are four such matrices. The other two are 
\be
\gamma_3\equiv \tau_y, \quad {\rm and} \quad \gamma_4 \equiv \tau_zs_y.
\label{two-g}
\ee
The Kane-Mele coupling term in (\ref{two-qubits}) is then associated with the matrix 
\be
\gamma_5\equiv - \gamma_1\gamma_3\gamma_2\gamma_4 = \tau_zs_z. 
\ee
Thus, all inter-qubit couplings  become linear when they are written in terms of Dirac $\gamma$-matrices. 
The commuting Hamiltonian $H'$ from (\ref{two-qubits2}) can also be naturally rewritten in terms of the $\gamma$-matrices because
\be
\tau_x s_z =i\gamma_3\gamma_5, \quad {\rm and} \quad s_x = i\gamma_4 \gamma_5.
\label{tow-g4}
\ee

Next, we recall that Dirac $\gamma$-matrices have a natural generalization to an arbitrary even spacial dimension.
 A  Wigner-Jordan  fermion representation (also known in the literature as the Wigner-Jordan transformation) of a  finite quantum spin chain with $j = 1, \ldots, N$ labeling the spins, can be written in a form
\begin{eqnarray}
\label{WJ-transform} \psi_{j} = \sigma_{j}^{x} \prod_{k=1}^{j-1} \sigma_{k}^{z}, \;\;\; \psi_{N+j} = \sigma_{j}^{y} \prod_{k=1}^{j-1} \sigma_{k}^{z},
\end{eqnarray}
 where we introduced operators that satisfy the fermion commutation relations
\begin{eqnarray}
\label{WJ-transform-2} \{ \psi_{j}, \psi_{k} \}= 2 \delta_{jk}, \; j = 1, \ldots, 2N.
\end{eqnarray}
 Alternatively, the operators $\psi_{j}$ can be viewed as $2^{N} \times 2^{N}$ matrices that represent the fermion operators acting in the $2^{N}$-dimensional space of $N$ spins. By introducing an additional operator
\begin{eqnarray}
\label{WJ-transform-3} \psi_{2N+1} = (-i)^{N} \prod_{j=1}^{N} \psi_{j}\psi_{j+1} = \prod_{j=1}^{N} \sigma_{j}^{z},
\end{eqnarray}
we obtain the anticommutation relations
\begin{eqnarray}
\label{WJ-transform-3} \{ \psi_{j}, \psi_{k} \} = 2 \delta_{jk}, \; j = 1, \ldots, 2N+1,
\end{eqnarray}
which is a $2^{N}$-dimensional representation of the Clifford algebra with $(2N+1)$ generators. On the other hand,
the corresponding $2^{N}$-dimensional Lie algebra representations can be integrated to the group representations known as spinor representations of ${\rm Spin} (2N)$ and ${\rm Spin} (2N+1)$ (see Ref.~\cite{kirillov}  for the  explicit construction of the spinor representations).

Spinor representations are used in quantum field theory to build relativistic fermions in higher-dimensional spaces, equipped with Euclidean signature, where the operators $\psi_{j}$ play the role of higher dimensional analogue of the Dirac matrices, and therefore will be hereafter denoted $\gamma_{j}$. We also define $\gamma_{\rm c} \equiv \gamma_{2N+1}$.

By collecting terms that have the same form as the terms in the pair of two-state Hamiltonians (\ref{two-qubits}) and (\ref{two-qubits2}) we  construct  the generalization of this pair to higher-dimensional Dirac $\gamma$-matrices: 
\begin{eqnarray}
\label{H-1-gamma} H_1(t, \varepsilon) &=& \varepsilon \gamma_{\rm c}  + \sum_{j=1}^N  \left(\beta_j t
i \gamma_{N+j}\gamma_{j} + g_j {\gamma_j} \right), \\
\label{H-2-gamma}
H_2 (t, \varepsilon) &=& t \gamma_{\rm c} \nonumber \\ &+& \sum_{j=1}^{N} (\beta_{j}^{-1} \varepsilon i \gamma_{N+j}\gamma_{j} + \beta_{j}^{-1} g_{j} i  \gamma_{N+j} \gamma_{\rm c}),
\end{eqnarray}
where  $\beta_i$, and $g_i$ are constant parameters; $\varepsilon$ is a constant in $H_1$ but is treated as the physical time in $H_2$.
Using the Clifford algebra relations, it is easy to check that the Hamiltonians $H_1$ and $H_2$ satisfy the integrability conditions (\ref{cond1})-(\ref{cond2}).


In order to return to  the Pauli operators $\sigma^{\alpha}_k$ of $N$ spins, where $\alpha\!=\!x,y,z$ and $k\!=\!1,\ldots, N$ is the spin's index, we define the coupling operators:
\begin{eqnarray}
\label{string1}
&&{\bf I}. \,\gamma_1\!=\! \sigma_{1}^{x}, \quad \gamma_2 \!=\! \sigma_{2}^{x} \sigma_{1}^{z}, \cdots, \quad \gamma_N \!=\! \sigma_{N}^{x} \prod \limits_{k=1}^{N-1} \sigma_{k}^{z}, \\
\label{string2}
 \nonumber &&{\bf II}. \, \tilde{\gamma}_1\!=\! \sigma_{1}^{x} \prod \limits_{k=2}^{N} \sigma_{k}^{z} ,\cdots,\,\,\,  \tilde{\gamma}_{N-1} \!=\!\sigma^{x}_{N-1} \sigma^{z}_{N},  \,\, \tilde{\gamma}_N \!=\!\sigma^{x}_{N}, 
\end{eqnarray}
which satisfy parafermion algebra relations  \cite{para}:
\begin{eqnarray}
\label{gamma-mt1}
\{ \gamma_i, \gamma_j \} &=&\{ \tilde{\gamma}_i, \tilde{\gamma}_j \}=2\delta_{ij},\\
\label{gamma-mt2}
[ \gamma_i, \tilde{\gamma}_j ] &=& 0,  \quad \forall \,i,j.
\end{eqnarray}
Then the two Hamiltonians have the form
\begin{eqnarray}
\label{ham-string1}
H_1(t,\varepsilon) &=&\varepsilon \prod_{j=1}^N \sigma_{j}^{z}  +\sum_{j=1}^N  \left( \beta_j t \sigma_{j}^{z} + g_j {\gamma_j} \right),\\
\label{ham-string2}
H_2 (t,\varepsilon) &=&t \prod_{j=1}^N \sigma_{j}^{z} +\sum_{j=1}^N  \left(\frac{ \varepsilon}{ \beta_j}  \sigma_{j}^{z} + \frac{g_j}{\beta_j} \tilde{\gamma_j} \right).
\end{eqnarray}
For the two-time vector
$$ (t, \varepsilon),$$
they satisfy (\ref{cond1}) and (\ref{cond2}). Hence they are integrable and belong to the two-time Landau-Zener family \cite{large-class}.  The example with $N=3$ shows that this family can be extended to add more parameters and independent commuting Hamiltonians but we will not do this because finding just one pair is sufficient to determine the transition probability matrix.

Since $H_1$ and $H_2$ have a particularly simple form when they are written in terms of the $2^N$-dimensional Dirac $\gamma$-matrices, and since these models describe interacting spins, we will call them  {\it $\gamma$-magnet} Hamiltonians. Let us focus on  
the $\gamma$-magnet $H_1$. It describes a heterogeneous qubit system because each qubit is coupled differently from any other qubit in the system. However, this system cannot be considered disordered because all interaction terms are taken to be $\gamma$-matrices that satisfy simple algebraic relations (\ref{gamma-mt1}). 

\begin{figure}
\scalebox{0.22}[0.22]{\includegraphics{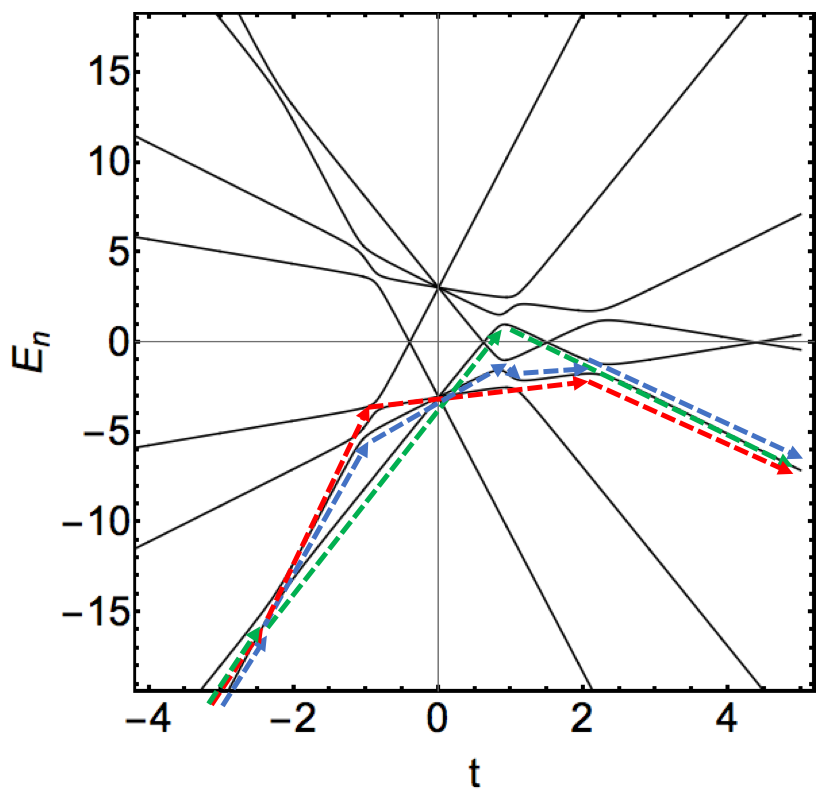}}
\caption{Time-dependent spectrum of the $\gamma$-magnet (\ref{ham-string1}) with $N\!=\!3$ interacting spins. The blue, green, and red arrows show three interfering semiclassical trajectories. The choice of parameters: $\varepsilon=3$, $\beta_1=2.83$, $\beta_2=1.35$, $\beta_3=3.5$,  $g_1=g_2=g_3=0.3$.}
\label{spectrum-16}
\end{figure}
In Fig.~\ref{spectrum-16}, we plot the energy spectrum of  $H_1(t)$ for different  $t$ and $N=3$. Integrability of this model can be inferred from the large number of points with  exact crossings of energy levels.
This feature is common for many  models of real nanomagnets \cite{garg1,garg2}.

Truly quantum behavior becomes evident  if we use the semiclassical approximation that can be justified for $\varepsilon \gg |g_i|$.
The diabatic states in the model (\ref{ham-string1}), i.e., the eigenstates of only the time-dependent part of $H_1(t)$,  are the spin projection states along $z$, e.g.,  the ground state at $t=-\infty$ is $|\ua \ua\ldots \ua \ra$ if $\beta_i>0\, \forall i$.
 According to the adiabatic theorem, all transitions between such states  are suppressed when energy levels are well separated from each other. This happens with the spectrum in Fig.~\ref{spectrum-16} as $t\rar \pm \infty$.

 However, inside the time interval shown in this figure, pairs of levels experience avoided crossings, i.e. the regions where levels do not cross exactly but appear very close to each other for  short time intervals. This happens when two diabatic energy levels (i.e. the eigenvalues of the time-dependent part of $H_1$) with a nonzero direct coupling between the corresponding diabatic states cross. For the spin with index $k$ in $H_1$, this can happen at
 \be
 t_{ k} = \pm \varepsilon/\beta_k,
 \label{resn}
 \ee
where the sign depends on the $z$-projections of all  spins.

After passing through such points, the system has finite amplitudes to stay on the initial level and to jump to a new one.  Thus, semiclassical trajectories can split from each other but then they also can merge by the end.  One can estimate the amplitude of a transition between any pair of states by summing amplitudes of all semiclassical trajectories that connect the initial state at $t=-\infty$ and the final state at $t=+\infty$.

 A common feature of all $\gamma$-magnets with $N>1$ is that there are generally more than one trajectory connecting pairs of different states.
An example is shown by red, green, and blue arrows in Fig.~\ref{spectrum-16}. All the marked trajectories  start from the ground state as $t\rar -\infty$, then split at different avoided crossings but then return to one energy level as $t\rar +\infty$. Such an interference is a signature of a nonclassical and many-body behavior. For example, it does not happen if spins are uncoupled from each other, so one can expect purely quantum mechanical effects to appear even for large $N$.

\section{Transition probabilities}
\begin{figure}[!htb]
\scalebox{0.33}[0.33]{\includegraphics{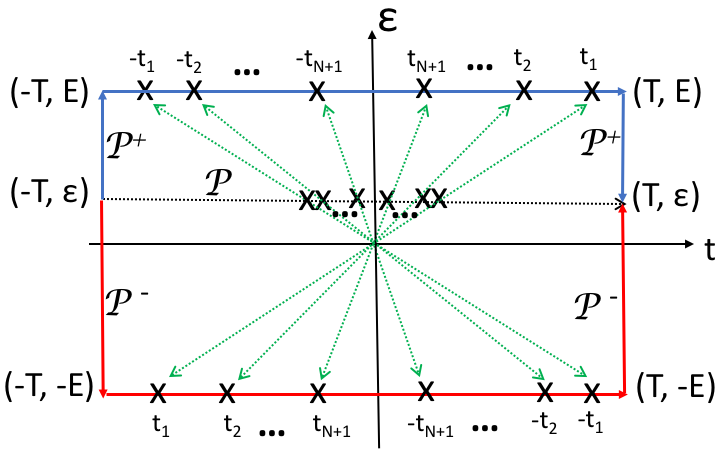}}
\caption{The physical evolution path ${\cal P}$ from $(-T,\varepsilon)$ to $(T,\varepsilon)$ (dotted black arrow) can be continuously deformed to either the path ${\cal P}^+$ (blue arrows) or the path ${\cal P}^-$ (red), where $E \gg \varepsilon$. Crosses ``X" mark positions of avoided crossings. Along ${\cal P}$ such points are close to each other, which leads to collective nonadiabatic dynamics. However, along ${\cal P}^+$ and ${\cal P}^-$, resonances are far apart. The chronological orders of resonances along ${\cal P}^+$ and ${\cal P}^-$ are opposite to each other. Green arrows mark positions of resonances during changes
of $\lambda=|E/\varepsilon|$.}
\label{paths}
\end{figure}

Let us now construct the matrix of transition probabilities between pairs of  diabatic states for evolution from $t=-\infty$ to $t=+\infty$.
Following \cite{commute}, we consider the evolution operator for a path ${\cal P}$ in the two-time space:
\be
\nonumber U=\hat{\cal T} {\rm exp}\left[-i\int_{\cal{P}} (H_1\, d t +H_2 \, d\varepsilon) \right],
\label{path1}
\ee
where $\hat{\cal T}$ is the path ordering operator.
Since we are interested in the effect of the sweep of the external field from large negative to large positive values,
the path ${\cal P}$ of our interest starts at time $-T$ and ends at $T$, where $T\rar \infty$. Along this path, we have $\varepsilon = {\rm const}$, as shown in Fig.~\ref{paths}. Let  $m$ and $n$ be the initial and the final diabatic states, respectively. Our goal is to find
the transition probabilities for all such pairs:
\be
P_{nm}=|U_{nm}|^2.
\label{transp1}
\ee

Due to (\ref{cond1})-(\ref{cond2}), $U$ does not depend on the choice of  the path $\cal{P}$, except the initial and final two-time points. This invariance follows from the fact that  the
  gauge field  $ {\bf A}=(-iH_1,-iH_2)$  has zero curvature. Hence,  ${\cal P}$ can be deformed to make either $|t|$ or $|\varepsilon|$ large \cite{commute}.
  In what follows, we change indices of spins so that
\be
\beta_1<\beta_2<\beta_3 < \ldots < \beta_{N},
\label{order-slopes1}
\ee
and we redefine the spin up and down states to make all $\beta_i$ positive.

Imagine now that we found the matrix of transition probabilities for any $H_1$ with $N$ spins. We add an extra spin to the model and redefine spin indexes to satisfy (\ref{order-slopes1}) again.
Consider the spin that has the largest slope: $\beta_{N+1}>\beta_N$.
The analysis then depends on whether the term $\varepsilon \prod_{i=1}^{N+1} \sigma^z_i$ in $H_1$ is positive or negative in the initial diabatic state at $t\!=\!-T$. First, let it be negative.
We then deform ${\cal P}$ into ${\cal P}^+$ as shown in Fig.~\ref{paths}, so that we initially change the Kane-Mele-like coupling from $\varepsilon$ to  $E=\lambda \varepsilon$, where $\lambda \gg 1$, 
then change time from $-T$ to $T$ and then return the coupling from $E$ to $\varepsilon$. At the left vertical leg of ${\cal P}^+$, we fix $t\!=\!- T$. Evolution along this leg is adiabatic because spin polarizations  are fixed in $H_2$ by infinitely large fields as  $T\!\rar \! \infty$. Hence, this leg only changes the phase of the initial state but does not lead to spin flips. The same is true for the right vertical leg of ${\cal P}^+$.

Resonances do happen along the horizontal part of ${\cal P}^+$ but, since $|E|\!\gg \!|g_i|$, all of them are now very well separated in time. Hence,
the semiclassical approximation that we already described can be applied and it becomes exact for $|E/g_i| \rar \infty$.
Then, among the resonances  (\ref{resn}), the first one that will happen will be at time
$$
t_{N+1}=|E|/\beta_{N+1}>0,
$$
at which only the spin-$(N+1)$ can flip. Since all other spins remain frozen during the passage through this resonance, the stay/flip probabilities for this spin are given by the Landau-Zener formula \cite{landau,zener,Majorana,stuckelberg}. Namely, let $p_{N+1}$ and $q_{N+1}$ be the probabilities for the $(N+1)$-st spin, respectively, to remain the same and to flip. Then
\be
p_{N+1}=e^{-\pi g^2_{N+1}/\beta_{N+1}}, \quad q_{N+1}=1-p_{N+1}.
\label{np1}
\ee
If this spin does not flip, then during the following evolution it cannot flip either because $t>\pm t_{N+1}$, and the
rest of spins flip as if they are uncoupled from this spin and have the initial condition $E' \prod_{i=1}^{N} \sigma^z_i <0$, where  $E'=E\sigma^z_{N+1}$, and where $\sigma^z_{N+1}$ is the polarization of  ($N+1$)st spin after it settles its value. Hence, the probability of any final diabatic state in this case will be $p_{N+1}$ times the probability of finding the given first $N$-spin configuration after evolution with the $N$-spin Hamiltonian, ignoring the spin $N+1$.

Alternatively, if the $(N+1)$-st spin flips, then it  creates the condition  $E' \prod_{i=1}^{N} \sigma^z_i >0$ for the rest of the spins. However,  all resonances at such a condition can happen only at negative times  $-t_k=-|E|/\beta_k$, while $t$ is already positive after passing through the resonance at $t_{N+1}$. Hence, in this case, none of the other spins will flip till the end of the evolution.

The case with initially $\varepsilon \prod_{i=1}^{N+1} \sigma^z_i>0$ becomes much more complex along ${\cal P}^+$ due to the path interference but it
 produces the same result because we have freedom to deform the path ${\cal P}$ into another path, ${\cal P}^-$, at which $\varepsilon \rar -\lambda \varepsilon$, where $\lambda \gg 1$, as shown in Fig.~\ref{paths}. At the beginning of the horizontal piece of this path we find that
$-\lambda \varepsilon \prod_{i=1}^{N+1} \sigma^z_i<0$, and the analysis reduces to the previous case.

To summarize, the transition probability matrix for the $\gamma$-magnet $H_1$, in which spin indices are changed to satisfy (\ref{order-slopes1}), can be constructed along a simple recursive process. Namely, consider a sequence of models (\ref{ham-string1}) with $N=1,2,\ldots$, such that $(N+1)$-st Hamiltonian is different from $N$-th one only by adding  terms with
$g_{N+1}$ and $\beta_{N+1}$ couplings, and adding a $\sigma^z_{N+1}$ factor to the product of operators with $\varepsilon$ coupling.
Let $P_{|i_{N} \ra \rar |j_N \ra}$ be the transition probability between states with indices $i$ and $j$ in the $N$-spin model and denote
 \be
 p_n \equiv e^{-\pi g_{n}^2/|\beta_{n}|}, \quad q_n =1-p_n, \quad n=1, 2,\ldots .
 \label{pq-n}
 \ee
 We will mark the states of the model with $N+1$ spins as
$|i_N \ua \ra$ and $|i_N  \da \ra$, where $i_N$ marks the states of the first $N$ spins as in the $N$-spin model. The transition probabilities in the $(N+1)$-spin system
are given by the following rules:

(i) The only nonzero probabilities of processes that flip the $(N+1)$-st spin are given by
\be
P_{|i_N \ua \ra \rar |i_N \da \ra} =P_{|i_N \da \ra \rar |i_N \ua \ra} = q_{N+1}.
\label{probq1}
\ee

(ii) The probabilities of transitions that do not lead to $(N+1)$-st spin flip are given by
\be
P_{|i_N \ua \ra \rar |j_N \ua \ra} =P_{|i_N \da \ra \rar |j_N \da \ra} = p_{N+1}P_{|i_{N} \ra \rar |j_N \ra}.
\label{probq1}
\ee

For example, $N=1$ is a single spin two-state LZ  model with the matrix of transition probabilities:
\be
P^{N=1}=\left( \begin{array}{cc}
p_1 & q_1 \\
q_1 & p_1
\end{array} \right).
\label{prob-sol1}
\ee
For $N=2$, we find the matrix (\ref{prob-sol2}).
Iterating (i)-(ii), we find that for the $N$-spin $\gamma$-magnet the  probabilities of transitions from any initial diabatic state are given explicitly by the following rules:

(a) the probability of not flipping any spin is $P_0 = \prod_{i=1}^N p_i$;

(b) the probability to flip only the $i$-th spin is $P_{i} = q_i\prod_{k=i+1}^N p_k$;

(c) the probability of  flipping more than one spin is zero.

In appendix, we show numerical tests that confirm  (a)-(c) up to $N=8$.

\section{Dynamic spin localization}
 The last property, (c), is the central result of this article. It defines the DSL behavior, i.e., that quantum many-body effects prevent propagation of spin-flips during time-dependent changes of parameters despite there are spin-flipping couplings for all spins. In fact, even the commutation of operators $H_1$ and $H_2$ is not a conservation law here due to the explicit time-dependence of parameters.  For large classical or quantum spin  systems, it is conceivable to construct a model with simple interactions that suppress multi-spin flips for some initial conditions during a chirp of a magnetic field. However, our model shows this behavior for arbitrary values of all parameters and arbitrary initial conditions. 

We are not aware of similar to DSL behavior in any known spin system. Among distantly related dynamic effects we mention that
there is evidence for slowing down heating of finite spin systems that are driven by  sufficiently weak periodic  pulses \cite{dle,dle-nanomag}. This is manifestation of, so-called, dynamic localization in  AC-field \cite{dunlap}. Another somewhat related effect is electron localization in energy space in a uniform electric field \cite{gefen-87}.    
However, DSL is essentially different, e.g., unlike the localization length in \cite{dunlap,gefen-87}, the maximal number of spin flips in our model does not depend on the driving field ramp.


There are totally $2^N$ diabatic states of $N$ spins.  Even if we adjust couplings to make all non-zero probabilities equal to $1/(N+1)$, we find that the final entropy
\be
S\equiv -\sum_{k=1}^{2^N} P_{n\rar k} \log_e P_{n\rar k},
\label{entropy}
\ee
for $\gamma$-magnets is saturated at
$
S_{max}^{\gamma}=\log_e (N+1).
$
 In contrast, for $N$ independent spins in a time-dependent field, $H=\sum_{i=1}^N[(\beta_it +\varepsilon_i)\sigma^z_i +g_i \sigma^x_i]$,  there is a finite probability to find any spin  configuration at the end.
The  entropy is then saturated when each spin has the flipping probability $1/2$:
$
S_{max}=N \log_e 2.
$
Thus, the final entropy of a $\gamma$-magnet per field sweep cannot increase by more than a value $\sim \log(N)$ versus $\sim N$ for noninteracting spins.

Property (b)  provides another practically interesting feature of $\gamma$-magnets. Imagine that $g_1\!\gg \!g_2 \!\gg \!\ldots \!\gg \!g_N$, whereas all
 $\beta_i$ are comparable. There is then a possibility to flip only  spin $k$ no matter what is the initial state by changing only one parameter. Namely, all $\beta_i$ are proportional to the external field sweeping rate, $d{B}/dt$, such as in Eq.~(\ref{two-qubits}). By changing  this rate, we rescale all $\beta_i$ by the same factor. We can then choose the sweeping rate so that for the indices $i$, $i>k$, the dynamics is strongly nonadiabatic, i.e., $p_i\rar 1$, whereas for the $k$-th spin it is adiabatic so that  $q_k\rar 1$. According to (b),  only the $k$-th spin will  then have  almost unit probability to flip. Thus, by varying only the external field ramp, we will be able to change the $z$-projection of any spin keeping other spins intact. This property is not  found in noninteracting spins.

\section{Discussion}

We constructed the $\gamma$-magnet Hamiltonian such that coupling terms that flip spins are essentially different for all spins. This Hamiltonian has considerable symmetry  when it is written in terms of Dirac $\gamma$-matrices. Due to such properties, $\gamma$-magnets do not fit  the types of commonly studied   interactions, e.g.,  spin glasses, regular spin lattices, and even commonly known spin cluster models, such as Gaudin magnets \cite{yuzbashyan}. 

We found a phenomenon of dynamic spin localization (DSL), which is  unusual both for  quantum and classical spin systems. It essentially relies on quantum interference, so it may be a purely quantum effect that has no classical counterpart.
 This suggests that heterogeneous interacting qubits may be a source of essentially novel physical phenomena and possibilities for quantum state control. 
Therefore, such interacting qubit systems  deserve attention on their own, without assuming that they  belong to the previously studied  types of spin matter.  

The transition probability matrix of any multistate Landau-Zener model with nondegenerate diabatic levels depends continuously on all couplings and level slopes. Hence, even if we add small interactions that break integrability then the effect of the time-dependent field remains almost the same, except for extremal values of some of the parameters. Therefore, there are {\it deffinitely} domains of parameters for which non-integrable interacting qubit systems show similar to DSL behavior. 

Other integrable many-body time-dependent models (for their reviews, see~\cite{yuzbashyan,quest-LZ})  show the behavior that is found in non-integrable systems too. For example, experimentally realized dynamic conversion of ultracold fermions into a molecular Bose condensate is well modeled by the integrable driven Tavis-Cummings model~\cite{DTCM,DTCM1,altland3}; and suppression of nonadiabatic excitations in the  time-dependent BCS Hamiltonian is also found in numerical simulations of more complex models~\cite{bcs}.

On experimental side, we hope that our work will stimulate interest in effective quibt Hamiltonians of localized states in Dirac materials. Additional degrees of freedom of Dirac electrons are often mentioned as an opportunity for quantum computation.
Within a localized state, such electrons can realize types of interactions that are generally needed to entangle qubits. We showed that, in addition, there can be a detailed  analytical understanding of such interactions, which can be used to design efficient control protocols.
Possibilities to create related integrable models are numerous and they still remain to be explored. In addition to a variety of such different models, the same $\gamma$-magnet Hamiltonians may have different physical interpretations.  For example, the representation of $\gamma$-matrices in (\ref{WJ-transform}) is not unique. For $N=3$, we can choose
\begin{eqnarray}
\nonumber \gamma_1&=& \tau_x\sigma_z, \quad \gamma_2=s_y\sigma_y, \quad \gamma_3=\sigma_x, \\
\gamma_4&=&\tau_y\sigma_z, \quad \gamma_5=s_x \sigma_y, \quad \gamma_6=s_z\sigma_y.
\label{gamma-new}
\ee
In terms of the pseudospins, the Hamiltonian (\ref{WJ-transform}) then reads
\begin{eqnarray}
\nonumber H_1 &=& \varepsilon \tau_z \sigma_z +t \left(\beta_1\tau_z +\beta_2s_z +\beta_3 s_z \sigma_z \right)+ \\
&+& g_1 \tau_x\sigma_z +g_2 s_y\sigma_y +g_3 \sigma_x.
\label{new-ham}
\ee
Note that this Hamiltonian includes only pairwise pseudospin interactions.
This Hamiltonian can describe two nearby zero energy localized states of a Dirac electron. The operator $\sigma_x$ then describes tunneling between these states,  the parameter $\beta_3$ follows from the difference of such state's spin $g$-factors, $g_2$ is the standard spin-orbit coupling effect on tunneling electrons \cite{four-LZ}, and $g_1$ describes mixing of states from different valleys; $\varepsilon$ may be unphysical but we can safely set it to zero.


It should be possible to study more complex $\gamma$-magnets experimentally too. 
Their Hamiltonians can be realized physically or programed in artificial qubit systems.
  Indeed, interactions in Eq.~(\ref{ham-string1}) are what is normally needed physically for implementing generalized quantum CNOT gates such that the state of a single qubit changes conditionally on states of several other qubits in the computational basis. Simple polynomial algorithms for implementing such gates using only the standard gate set exist \cite{tofoli-gate}. Another possibility  is to use the fact that interactions involving products of multiple qubit projection operators along $z$-axis can be  realized using only pairwise qubit couplings by employing additional qubits~\cite{squid1,squid2}.


It is unclear whether DSL-like behavior can be found in natural large spin systems. At this stage, we can only speculate. The first candidate is the class of molecular nanomagnets. 
Indeed, for $N=2$ the $\gamma$-magnet Hamiltonian has the same matrix form as a spin-3/2 nanomagnet~\cite{li-dynamic}, such as 
the molecule $V_{15}$ \cite{nagaosa-LZ}. Flexibility of nanomagnet synthesis  enables the design of spin systems with desirable properties:  strength and type of interactions,   long quantum coherence, control by means of optics, voltage, and magnetic fields~\cite{nanomag-rev,mol-coh, mol-coh1}. Nanomagnets are already  used as quantum information hardware  \cite{mol-qc1,mol-qc2}.
For nanomagnets, macroscopic quantum tunneling in multi-spin configuration space is observable  during  linear-in-time changes of the magnetic field as a staircase of magnetization steps.
Interference between  tunneling pathways  is then found  as the suppression of some of such steps~\cite{mol-interf,mol-interf1,mol-interf2,mol-interf3}.

However, spins in nanomagnets are normally coupled by exchange interactions. Our discussion of Dirac Hamiltonians suggests that DSL can be more likely found when spins interact via the spin-orbit coupling. Hence, DSL-like behavior may be found in nanomagnets with strong \cite{large-DM} or artificially enhanced \cite{balk-DM} Dzyaloshinskii-Moriya interactions. 

It is also interesting to find connections between DSL and many-body localization \cite{mbl1,mbl2} in strongly disordered spin systems. There are important differences: DSL occurs in multi-spin phase space and does not require disorder. Nevertheless, the many-body localization emerges when a spin system behaves locally as an integrable model with complex  interactions \cite{mbl2} that has a large but essentially finite number of commuting operators. If this property is preserved in a wide range of strong external magnetic fields, then the time-dependent integrability may also emerge when this field becomes time-dependent. Therefore, the many-body localized spin systems are potential candidates  for finding DSL  in quantum materials. 

Finally, there are many similarities between our heterogeneous qubits and qubit Hamiltonians that are used in quantum annealing  computations. 
The latter  realize quantum evolution with explicit time dependence:
\be
\hat{H}_{qa}(t) = \hat{H}_A(\hat{s}^z_1,\ldots, \hat{s}^z_{N}) +g(t)\hat{H}_B(\hat{\bm s}_1,\ldots, \hat{\bm s}_{N}),
\label{aham1-11}
\ee
where $\hat{H}_A$ contains only Ising-like qubit couplings and $\hat{H}_B$ has  a ground state that is relatively easy to prepare. 
 The parameter $g(t)$ is initially  taken to be large enough to make the second term in (\ref{aham1-11}) completely dominate $\hat{H}_A$, but $g(t)$  later decays to zero  as $t\rar \infty$. According to the adiabatic theorem,  a very slow decay of $g(t)$ converts the  initial ground state  (of $\hat{H}_B$) to the  final ground state {(of $\hat{H}_A$)}, which is then read by measuring spins along the $z$-axis.  
 
 
A classical optimization problem encoded in the Ising part, $\hat{H}_A$, introduces couplings that are neither regular nor taken from some distribution. Each Ising spin plays a specific role in this problem and has unique for it set of links to other spins. In this sense, $\hat{H}_A$ resembles the $\gamma$-magnet. 
DSL-like behavior  can be a considerable problem for quantum annealing computations because DSL prevents a spin system from exploring its phase space during time-dependent parameter driving. This analogy suggests that DSL can be searched using quantum annealing machines by exploring  the problems that produce anomalously large amounts of computational errors.

\appendix
\begin{widetext}
\section{Numerical test of the solution for the $\gamma$-magnet}

We found the transition probabilities for the $\gamma$-magnet with the Hamiltonian (6)  up to  $N=8$ numerically 
by discretizing the evolution in small time steps, $\Psi(t+dt) =U(t,dt) \Psi(t)$ where $U(t,dt) =e^{-i H(t) dt}$,  and finding the matrix exponent for each step.
\begin{figure}
\scalebox{0.35}[0.35]{\includegraphics{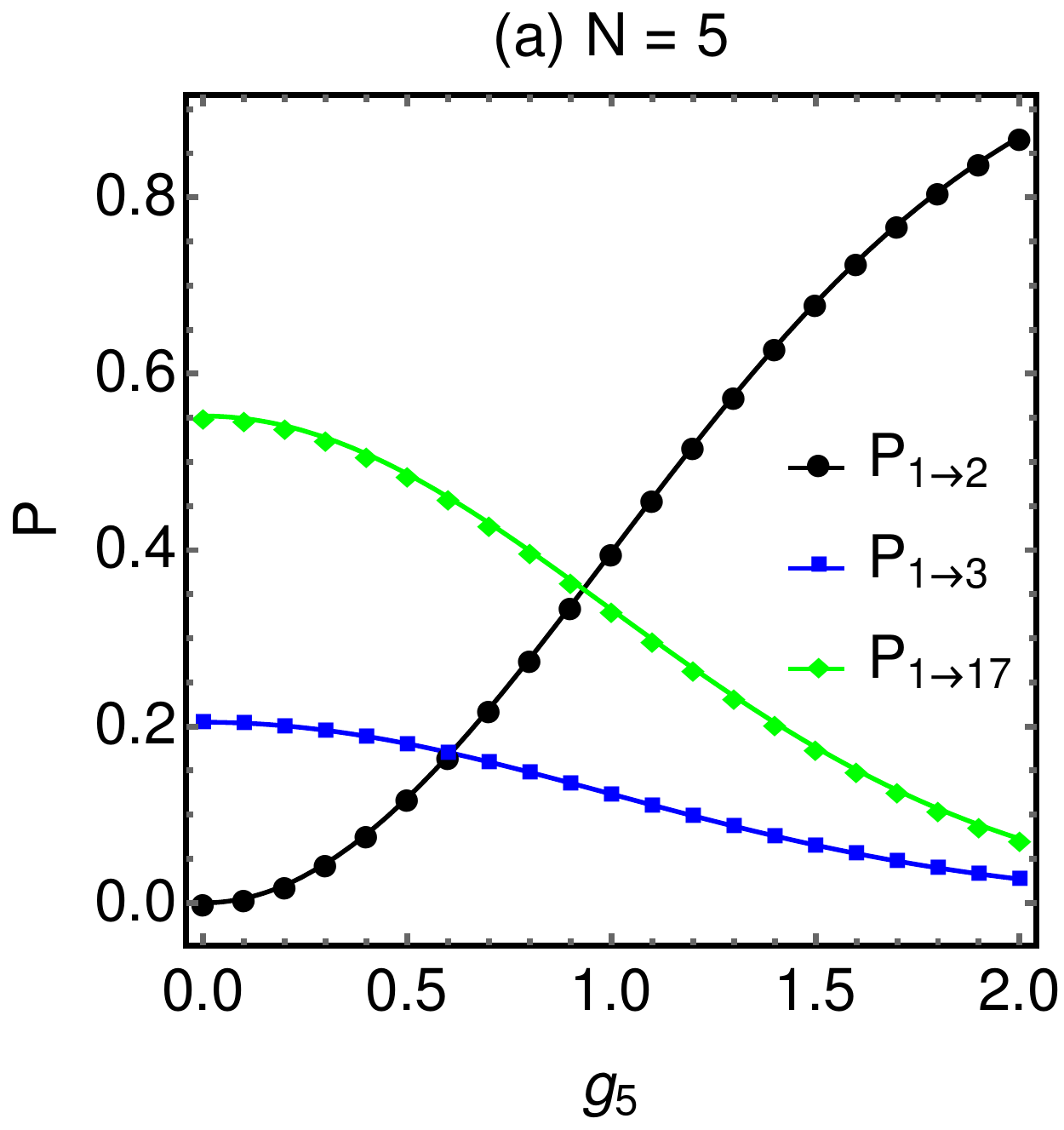}}
\scalebox{0.36}[0.36]{\includegraphics{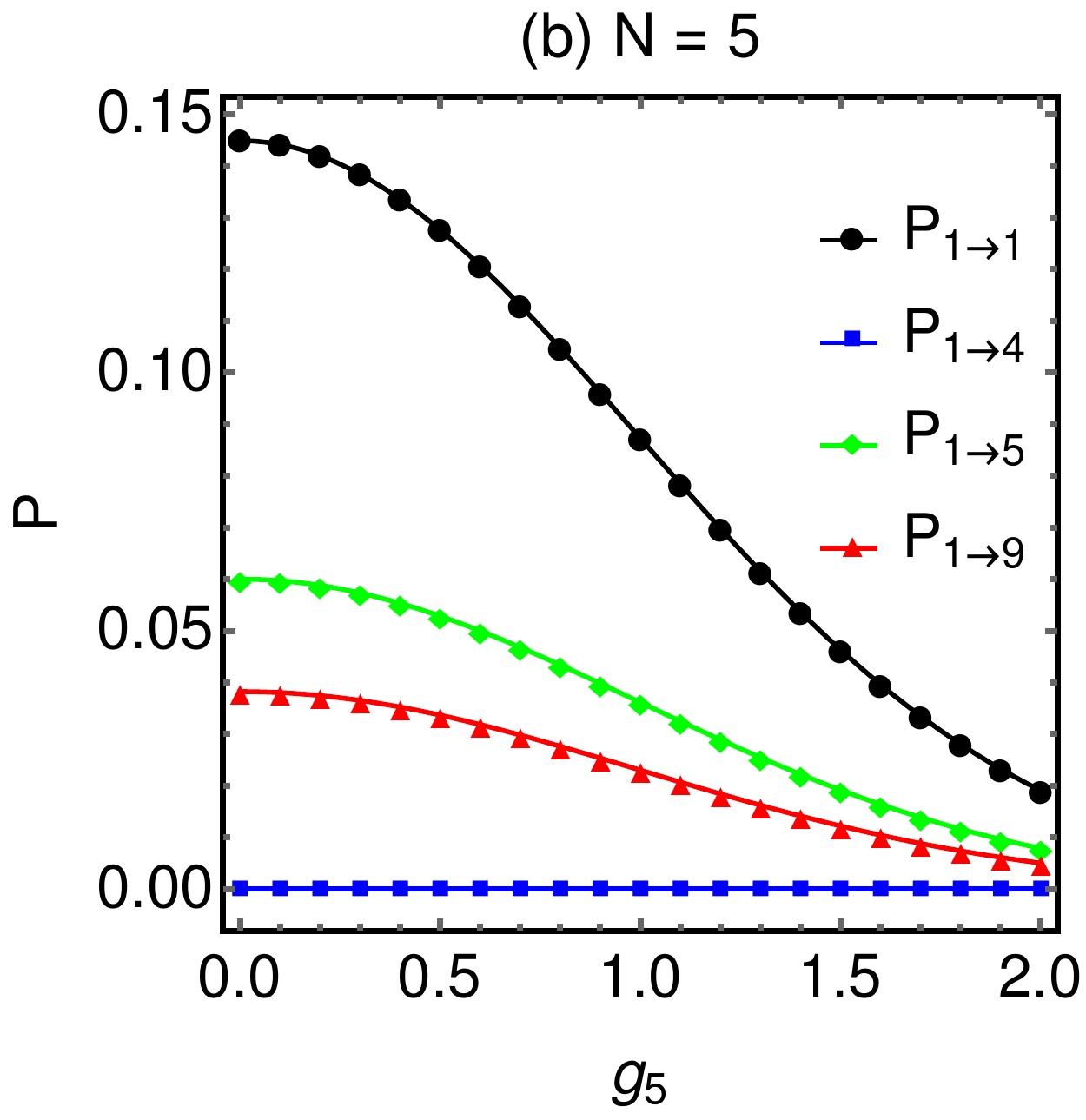}}
\scalebox{0.35}[0.35]{\includegraphics{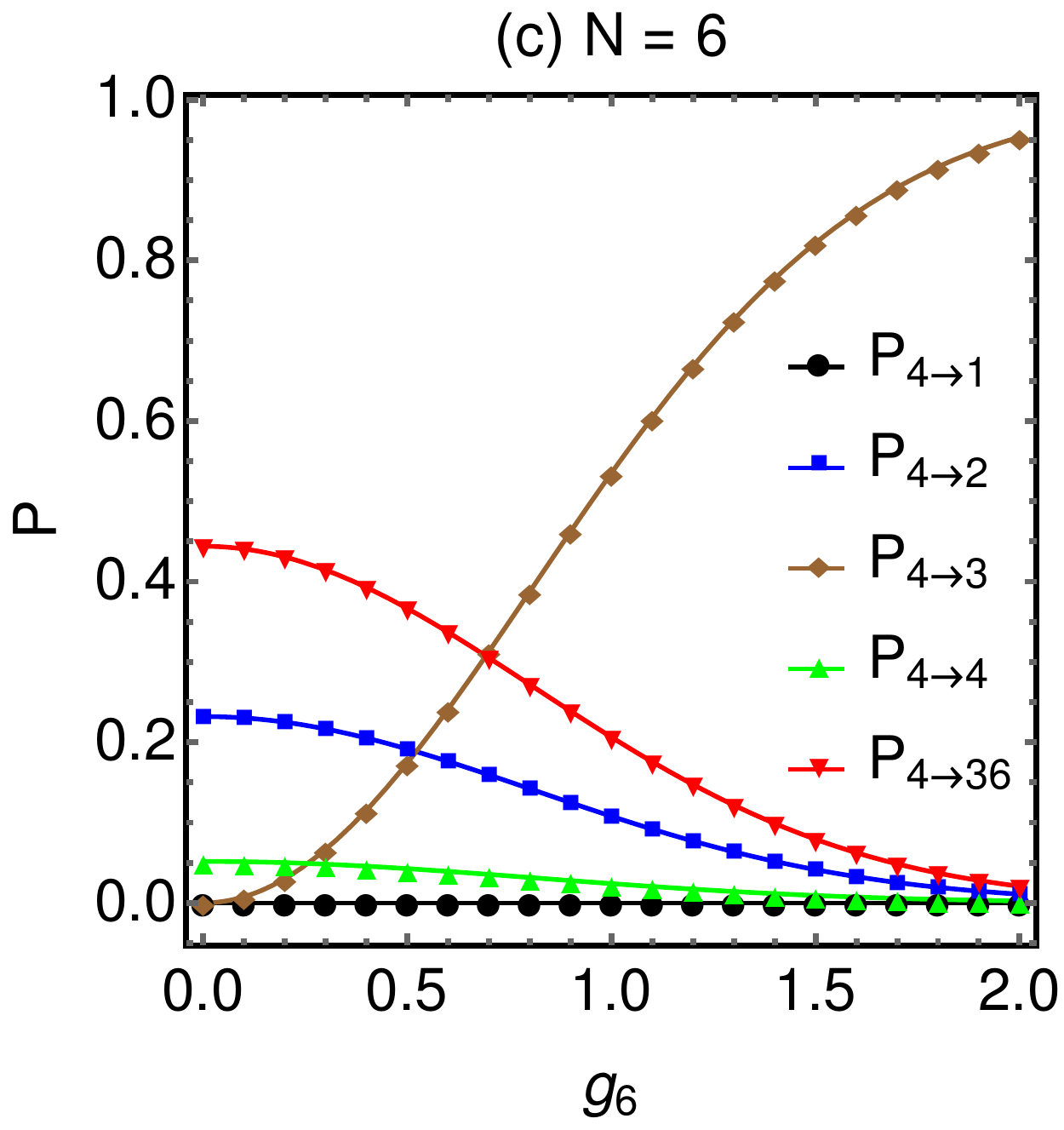}}\\
\scalebox{0.35}[0.35]{\includegraphics{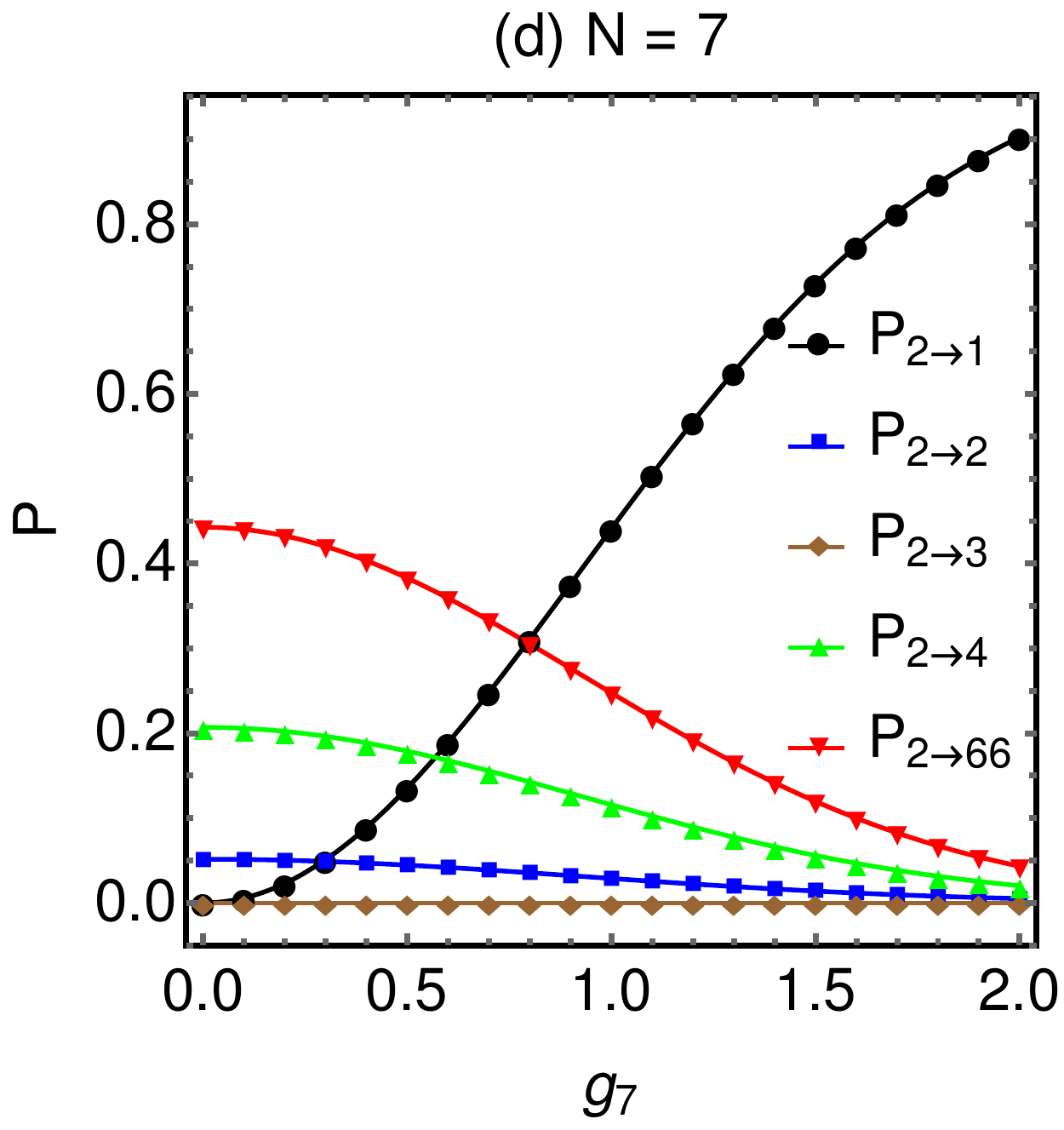}}
\scalebox{0.35}[0.35]{\includegraphics{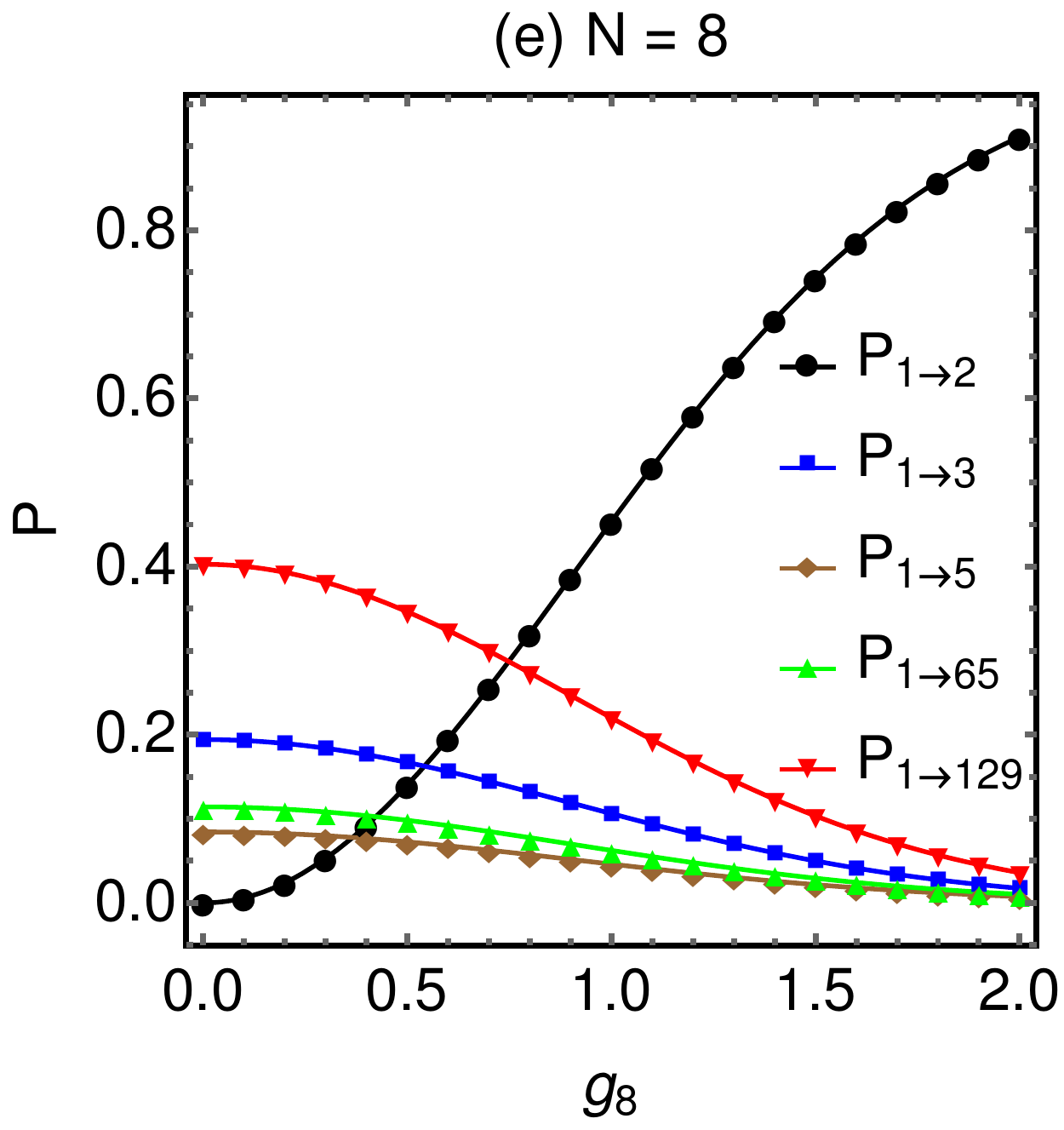}}
\scalebox{0.36}[0.36]{\includegraphics{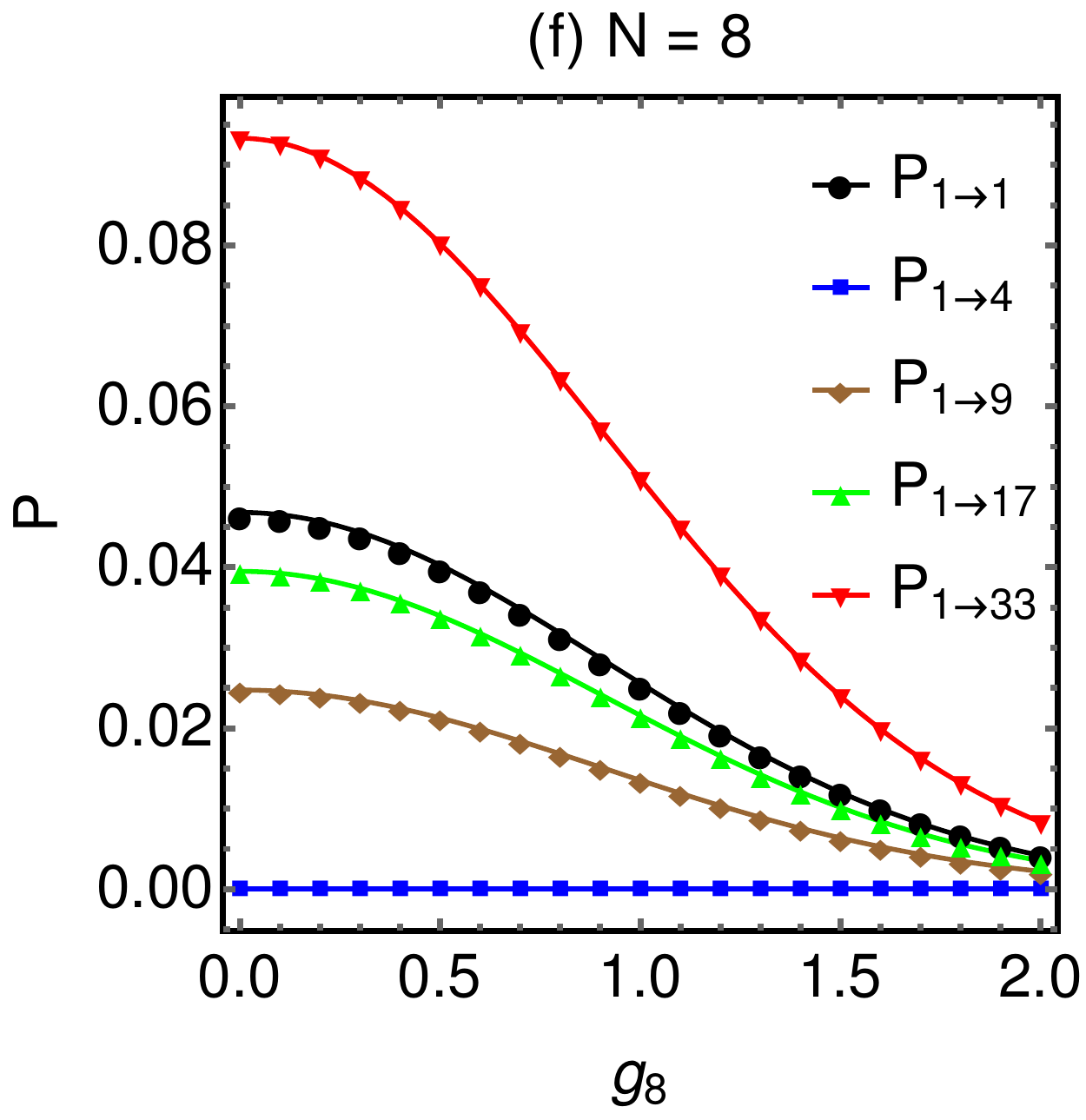}}
\caption{The transition probabilities for the $\gamma$-magnets with: (a) and (b) $N=5$, (c) $N=6$, (d) $N=7$, (e) and (f) $N=8$. Points are results of numerical simulations and solid curves are analytical predictions. The labeling of states is such that the state $| \sigma_1, \sigma_2,...\sigma_N \ra$ with $\sigma_i=0$ for spin $\ua$ and $\sigma_i=1$ for spin $\da$ is labelled by the decimal number converted from the binary number ``$\sigma_1 \sigma_2...\sigma_N$'' plus $1$. For example, the spin configurations appeared in (a) and (b) are: $|1\ra=|\ua\ua\ua\ua\ua\ra$, $|2\ra=|\ua\ua\ua\ua\da\ra$, $|3\ra=|\ua\ua\ua\da\ua\ra$, $|4\ra=|\ua\ua\ua\da\da\ra$, $|5\ra=|\ua\ua\da\ua\ua\ra$, $|9\ra=|\ua\da\ua\ua\ua\ra$, $|17\ra=|\da\ua\ua\ua\ua\ra$. 
Note that the probability $P_{1\rar 4}$ is zero because states $|1\ra$ and $|4\ra$ are different by the direction of a pair of spins.
The spin configurations appeared in (c) are: $|1\ra=|\ua\ua\ua\ua\ua\ua\ra$, $|2\ra=|\ua\ua\ua\ua\ua\da\ra$, $|3\ra=|\ua\ua\ua\ua\da\ua\ra$, $|4\ra=|\ua\ua\ua\ua\da\da\ra$, $|36\ra=|\da\ua\ua\ua\da\da\ra$. Parameters: (a) and (b) $\varepsilon = 1$, $\beta_1 = 0.5$, $\beta_2 = 1.7$, $\beta_3 = 4.1$, $\beta_4 = 5.1$, $\beta_5 = 6.2$, $g_1 = 0.5$, $g_2 = 0.17$, $g_3 = 0.32$, $g_4 = 0.61$, and $g_5$ changes from $0$ to $2$; (c) $\varepsilon =1$, $\beta_1=0.5$, $\beta_2=1.7$, $\beta_3=2.1$, $\beta_4=3.1$, $\beta_5=3.6$, $\beta_6=4.1$, $g_1=0.6$ $g_2=0.35$, $g_3=0.32$, $g_4=0.24$, $g_5=0.55$, and $g_6$ changes from $0$ to $2$; (d) $\varepsilon =1$, $\beta_1=0.5$, $\beta_2=1.7$, $\beta_3=2.1$, $\beta_4=3.1$; $\beta_5=3.6$, $\beta_6=4.1$, $\beta_7=5.4$, $g_1=0.6$, $g_2=0.35$, $g_3=0.32$, $g_4=0.24$, $g_5=0.2$, $g_6=0.55$, and $g_7$ changes from $0$ to $2$; (e) and (f) $\varepsilon = 1$, $\beta_1 = 0.5$, $\beta_2 = 1.7$, $\beta_3 = 2.1$, $\beta_4 = 3.1$, $\beta_5 = 3.6$, $\beta_6 = 4.1$, $\beta_7 = 4.4$, $\beta_8 = 5.2$, $g_1 = 0.6$, $g_2 = 0.35$, $g_3 = 0.32$, $g_4 = 0.24$, $g_5 = 0.2$,  $g_6 = 0.38$, $g_7 = 0.55$, and $g_8$ changes from $0$ to $2$. }
\label{gamma-magnet-N=5678}
\end{figure}
In Fig.~\ref{gamma-magnet-N=5678}, we show  the transition probabilities for $N=5$, $6$, $7$ and $8$, and various initial conditions. Only some of the transition probabilities are shown.  The corresponding theory predictions by the rules (a) - (c)  from the main text  are marked by the solid curves. For example, for $N=5$ the transition probabilities shown on the plots are:
\begin{align}\label{}
&|\ua\ua\ua\ua\ua\ra\rar|\ua\ua\ua\ua\ua\ra: P_{1\rar 1} =p_1 p_2 p_3 p_4 p_5,\nn\\
&|\ua\ua\ua\ua\ua\ra\rar|\ua\ua\ua\ua\ua\ra: P_{1\rar 2} = q_5,\nn\\
&|\ua\ua\ua\ua\ua\ra\rar|\ua\ua\ua\da\ua\ra: P_{1\rar 3} = q_4 p_5, \nn\\
&|\ua\ua\ua\ua\ua\ra\rar|\ua\ua\ua\da\da\ra: P_{1\rar 4}=0, \nn\\
&|\ua\ua\ua\ua\ua\ra\rar|\ua\ua\da\ua\ua\ra: P_{1\rar 5} =q_3 p_4 p_5,\nn\\
&|\ua\ua\ua\ua\ua\ra\rar|\ua\da\ua\ua\ua\ra: P_{1\rar 9}=q_2 p_3 p_4 p_5,\nn\\
&|\ua\ua\ua\ua\ua\ra\rar|\da\ua\ua\ua\ua\ra: P_{1\rar 17} = q_1 p_2 p_3  p_4 p_5,
\end{align}
with $p_n$ and $q_n$ defined in Eq.~(13) in the main text. All numerical results agree with the analytical predictions. Note, e.g., that states $|1\ra$ and $|4\ra$ here differ by the direction of a pair of spins, so this probability is identically zero. 
\end{widetext}

\acknowledgments
We thank A. Saxena for useful discussions. 
This work was supported by the U.S. Department of Energy, Office of Science, Basic Energy
Sciences, Materials Sciences and Engineering Division, Condensed Matter Theory Program (V.Y. C. and N.A.S.), and by the J. Michael Kosterlitz Postdoctoral Fellowship at Brown University (C.S.).
N.A.S. also thanks the support from the LDRD program at LANL. 

Authors made equal contributions to this article.




\end{document}